# First direct evidence of two stages in free recall and three corresponding estimates of working memory capacity

Eugen Tarnow, Ph.D.[1]

18-11 Radburn Road, Fair Lawn, NJ 07410, USA

etarnow@avabiz.com

**Abstract**

I find that exactly two stages can be seen directly in sequential free recall distributions. These distributions show that the first three recalls come from the emptying of working memory, recalls 6 and above come from a second stage and the 4th and 5th recalls are mixtures of the two.

A discontinuity, a rounded step function, is shown to exist in the fitted linear slope of the recall distributions as the recall shifts from the emptying of working memory (positive slope) to the second stage (negative slope). The discontinuity leads to a first estimate of the capacity of working memory at 4-4.5 items. Working memory accounts for the recency effect. The primacy effect comes from the second stage with a contribution also from working memory for short lists (the first item). The different slopes of the working memory and secondary stages, and that the two have different functional forms, accounts for the u-shaped serial position curve.

The total recall is shown to be a linear combination of the content of working memory and items recalled in the second stage with 3.0-3.9 items coming from working memory, a second estimate of the capacity of working memory.

A third, separate upper limit on the capacity of working memory is found (3.06 items), corresponding to the requirement that the content of working memory cannot exceed the total recall, item by item. This third limit presumably corresponds to the least chunked item. This is the best limit on the capacity of unchunked working memory.

---

[1] The author is an independent researcher. He received a PhD in physics from MIT. His research areas include semiconductor physics, marketing science, management science, and the psychology of dreams, conformity, obedience and short term memory.

**Introduction**

Free recall stands out as one of the great unsolved mysteries of modern psychology (Hintzman; 2011, for reviews, please see, for example, Watkins, 1974; Murdock, 1975; Laming, 2010; some believe existing computer models provide important insights, I do not, see Tarnow, 2012 for one set of computer models that cannot be believed).  Items in a list are displayed or read to subjects who are then asked to retrieve the items. It is one of the simplest ways to probe short term memory.  The results (Murdock, 1960; Murdock, 1962; Murdock, 1974) have defied explanation. Why do we remember primarily items in the beginning and in the end of the list, but not items in the middle, creating the famous u-shaped curve of probability of recall versus serial position?  Why can we remember 50-100 items in cued recall but only 6-8 items in free recall?  If one may speak about a consensus in memory psychology, that short-term memory has a limited capacity store typically named "working memory", where can this store be seen in free recall and what is its capacity?  These two latter issues will be covered in this article and the results will also include the origin of the u-shaped curve.

The effort to calculate the capacity of what was then termed "primary memory" using free recall was reviewed by Watkins (1974).  These calculations were based on indirect evidence of a division of memory and sometimes *a priori* assumptions were used.  The capacity was generally found to be three items. This was long before recent experiments finding a limit of working memory of exactly three items (Chen and Cowan, 2009; Anderson, Vogel and Awh, 2011; for an opposing view see Bays and Husain, 2008; and Bays, Catalao and Husain, 2009; for a review see Cowan, 2001).

In the Watkins (1974) review, "primary memory" is at the beginning defined narrowly as "the mechanism underlying the recency effect in free recall".  This is an unfortunate assumption because it prevents models from giving *ab-initio* pictures of working memory content.  He notes that the recency effect is relatively constant against experimental manipulation of list length, presentation rate, frequency of list items occurring in the language, semantic association, phonological similarity and concurrent activity. Several methods of measuring the capacity of primary memory were discussed, some of which calculate a capacity of about 3 items, all of which are problematic.

First, Waugh and Norman (1965) assumed that an item can be in working memory or in secondary memory, that the two representations are independent of each other, that the probability of being in secondary memory is constant for all items and that its value is the average of the middle item recall probabilities, and that if an item is in either store the chance is 100% it will be recalled.  Two problems with this model is that cueing of one item with another is explicitly disallowed and that secondary memory presumably contains much more than is found in free recall (as seen in recognition experiments). Watkins (1974) obtains a primary memory capacity using the Waugh and Norman method (as well as of a variation) which ranges from 2.21-3.43 words, presumably including cancellations of systematic errors. Similar estimates using this method were arrived at by Glanzer & Razel (1974). The model is *ab-initio* with respect to the capacity but not with respect to the item content of working memory.

Second, Murdock (1967) noted that total recall of long lists was a linear function of the presentation time. He proposed that the intercept was the capacity of working memory and that the linear rise was due to information processing. This *ab-initio* approach to the capacity does not limit the definition of primary memory to the recency effect (also emphasized by Watkins, 1974). One problem is information processing includes not just secondary memory items but also includes processing of primary items which need a minimum presentation time to be activated (see, for example, Tarnow, 2008), blurring the distinction between primary and secondary memory. The size of primary memory calculated is 6.1 (Murdock, 1960). Craik (1968) used the same method and a variation to arrive at working memory capacities of 4.9 words and 1.7 digits (the capacity for counties and animals were 2.1 and 3.5 items, respectively). A modification suggested by Legge rephrases this method to define primary memory capacity as the longest list length for which recall is perfect (cited in Craik, 1968). Legge's approach does not consider the possibility of cuing of items, potentially overestimating working memory. The model is *ab-initio* with respect to the capacity but does not predict the item content of working memory.

Third, Tulving & Patterson (1968) assumed that the last four items are the only ones that can be present in primary memory and the capacity of primary memory is the sum of probabilities of remembering the last four presented items. This approach limits *a priori* the capacity of working memory to four items, it discounts the possibility that individual rehearsal algorithms can differ, it contradicts the clear presence of other items in the first recall (the first recall should reflect only working memory), and it does not take into account the possibility that the total recall of the last four items includes recall from secondary memory. In a comparison of the different methods, Watkins (1974) obtains a primary memory capacity which ranges from 2.56-3.23 words, presumably including cancellations of systematic errors. The model is not *ab-initio*.

Fourth, Tulving & Colotla (1970) defined the capacity of primary memory as the sum of probabilities of recalling items with "intratrial retention interval" (the number of other items presented or recalled between the item presentation and recall) of 7 or less (Tulving & Colotla, 1970). This approach sets an *a priori* limit on the intratrial retention interval, thereby indirectly fixing the capacity of working memory. It also, presumably incorrectly, disallows first item recalls, discounts early recalls which include both the first and last items, and discounts recalls of the recency items depending upon the order of those recalls. Watkins (1974) obtains a primary memory capacity using this method which ranges from 2.93-3.35 words, presumably including cancellations of systematic errors. The model is not *ab-initio* with respect to either capacity nor the item content of working memory.

Watkins (1974) noted, for the methods he reviewed, that there is not necessarily an agreement between methods or even within the methods as to which items are in primary memory and which are not. Within the same method there is sometimes no consistency of the ratio of secondary to primary memory when varying list lengths, presentation rates and number of languages per list (in particular for the Waugh & Norman, 1965, method and its variation and for the Murdock, 1967, method).

Glanzer (1982) calculated the capacity of primary memory using the difference of total recall curves with and without a delay during which subjects were asked to count out numbers starting with a given number (data from Glanzer & Cunitz, 1966) and found that the capacity was 3. The differences in recall distributions in Glanzer & Cunitz indicated that primary memory is responsible for the recency effect. The measurement is *ab initio* as is the localization of primary memory though in both cases the evidence is indirect.

In this contribution I will show that exactly two stages can be seen *directly* in free recall distributions. The first stage is the emptying of working memory and capacity of working memory can be calculated *ab-initio* and is 3 for unchunked items and 3-4.5 for chunked items. The emptying of working memory is responsible for recency. The second stage is responsible for primacy; with working memory also contributing the first item for short lists.

**Method**

This article makes use of the Murdock (1962) and Murdock & Okada (1970) data sets (downloaded from the Computational Memory Lab at the University of Pennsylvania (http://memory.psych.upenn.edu/DataArchive). The former has been described as "one of the most comprehensive datasets ever collected" (Unsworth, Brewers & Spillers, 2011). There were six experiments typically labeled N-M where N is the total number of words presented and M is the intervals in seconds between word presentations. In Table 1 is summarized the experimental processes which generated the data sets used in this paper.

TABLES

| Work | Item types | List length and presentation interval | Recall interval (all were immediate) | Subjects | Item presentation mode |
|---|---|---|---|---|---|
| *Murdock (1962)* | *Selection from 4000 most common English words, referred to as the Toronto Word Pool.* | *10,15,20 words in a list each word presented every 2 seconds*<br><br>*20 ,30, and 40 words in a list, each word presented once a second* | 1.5 minutes | 103 undergraduates | Verbal |
| *Murdock & Okada (1970)* | Toronto word pool (1150 of the 4000 most common English words which have two syllables words not more than eight letters long with homophones, proper nouns, contractions, and archaic words deleted.) | 20<br><br>One and two words per second | 1.5 minutes | 72 undergraduate students from larger low level psychology courses | Visual |

*Table 1. Information about experiments included in the study.*

**Results & Discussion**

In Fig. 1a (Fig. 1b) is shown the recall distributions of recalls 1-8 from the 10-2 (40-1) dataset of Murdock (1962). These distributions show direct evidence for a two stage process. By definition the first recall comes from working memory, and from the similarity of the $2^{nd}$ and $3^{rd}$ recalls these also come from working memory. The last three recalls all look the same and come from a second stage. Recalls 4 and 5 are a combination of the two. In each recall is plotted a best linear fit which expresses the balance between recency (positive slope) and primacy (negative slope). As we see the slopes go from primacy for the emptying of working memory to recency for the second stage. Working memory can be seen as responsible for recency (consistent with previous work, see Watkins, 1974 and Glanzer, 1982); primacy comes from the secondary process though working memory adds the first items in the shorter 10-2 list; together they create a u-shaped serial position curve.

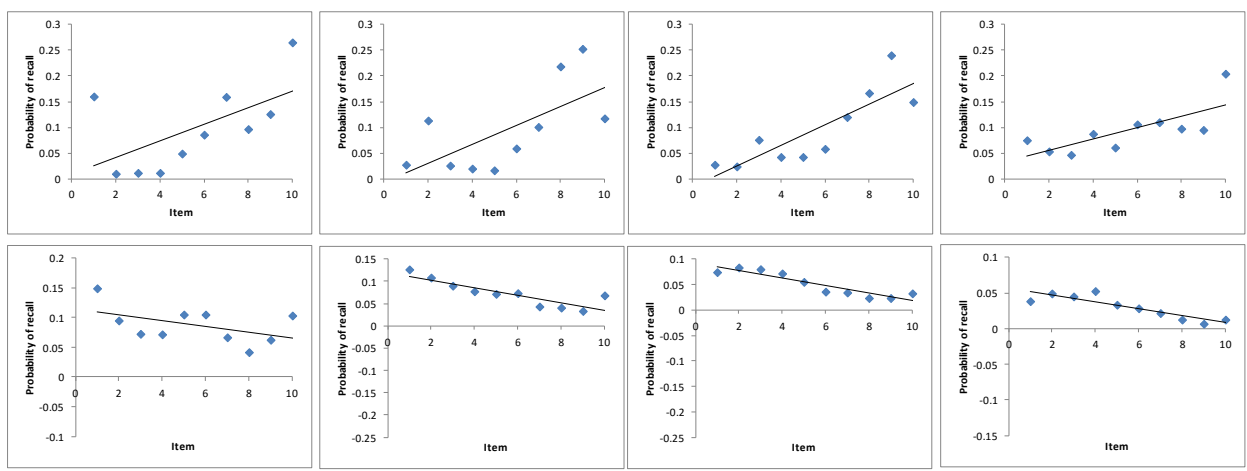

*Fig. 1a. Recalls 1-4 (top panel) and 5-8 (bottom panel) for the 10-2 experiment. The first three recalls are from working memory, last three recalls from second stage recall, and the $4^{th}$ and $5^{th}$ recalls are from a combination of working memory and second stage recall.*

Running head: First direct evidence of two stages in free recall

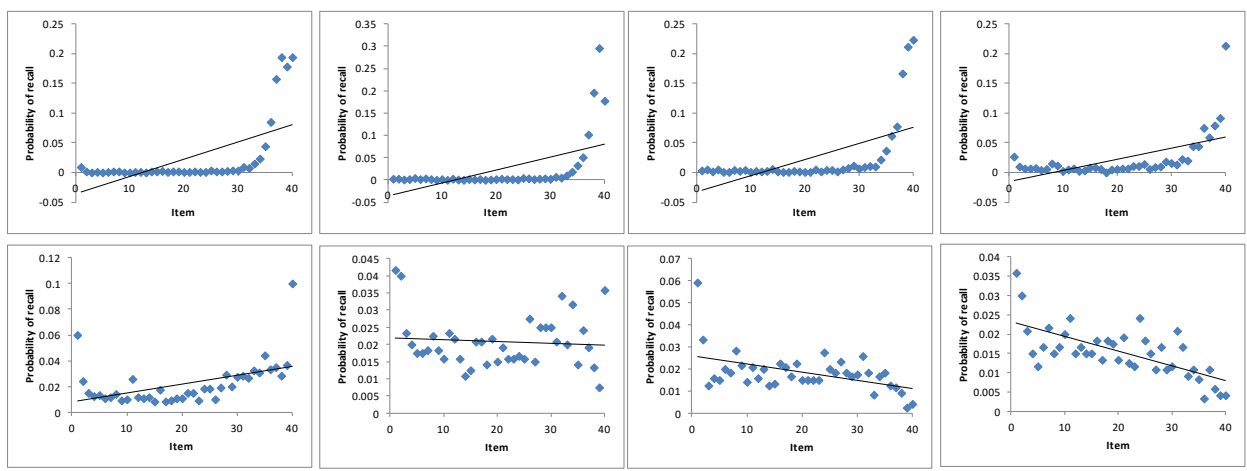

*Fig. 1b. Recalls 1-4 (top panel) and 5-8 (bottom panel) for the 40-1 experiment. The first three recalls are from working memory, last three recalls from second stage recall, and the 4th and 5th recalls are from a combination of working memory and second stage recall.*

The slopes as a function of recall are plotted in Fig. 2. The curve is a smoothed step function, the first discontinuity to be found to separate the emptying of working memory from the second stage. The middle of the discontinuity, which corresponds to the capacity of working memory, is 4 for the 10-2 data and 4.5 for the 40-1 data.

Running head: First direct evidence of two stages in free recall

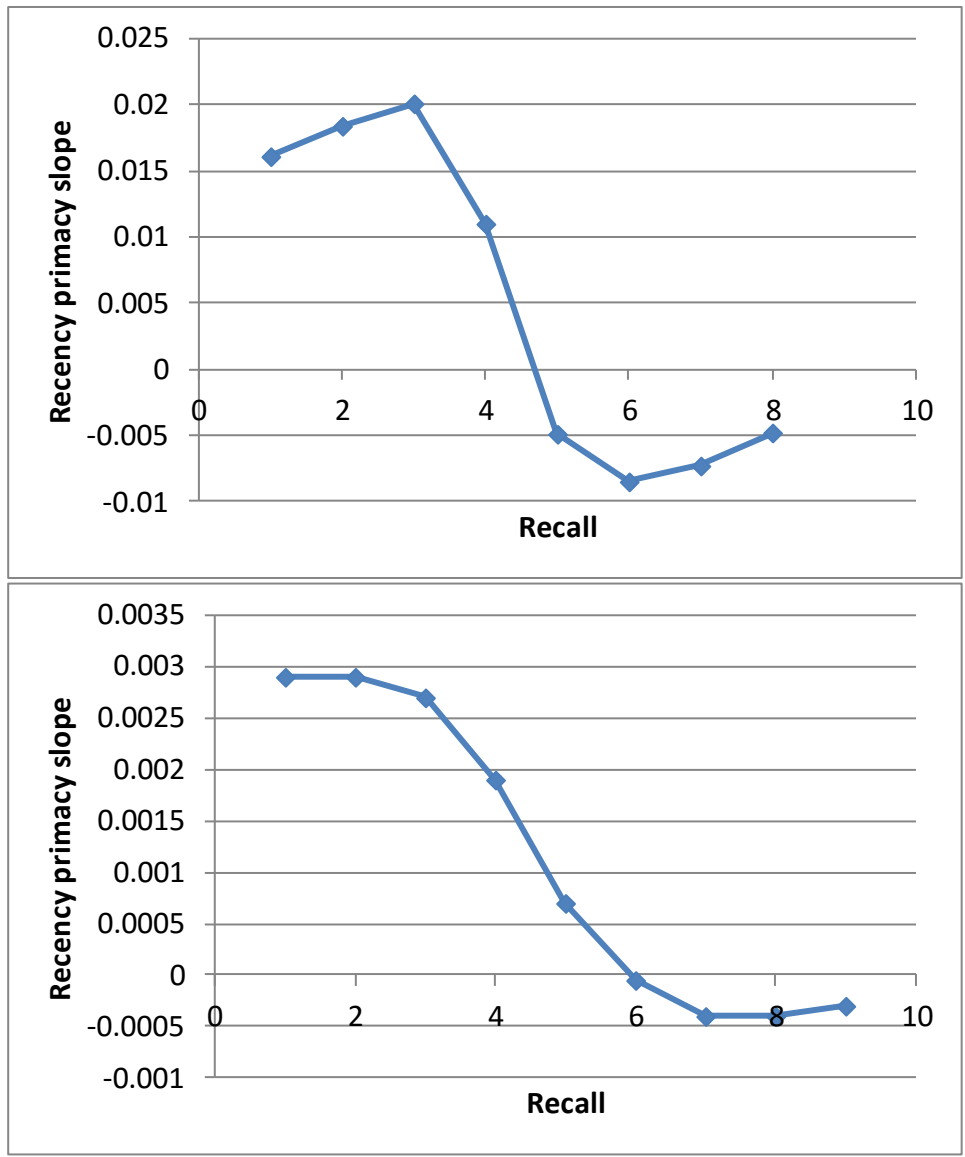

*Fig. 2. The slope of a linear fit to the serial position curves for the 10-2 (upper panel) and 40-1 (lower panel) data. Positive slope indicates recency, negative slope indicates primacy. Note the similarity to a step function. The middle of the step function is a little higher than 4, corresponding to the capacity of working memory.*

The boundary between the two stages is illustrated not only by the change from the emptying of working memory to the second stage in the recall distributions of Fig. 1a and 1b, but also by the first recall itself. In it some of the word recall probabilities are amplified compared to the overall recall distribution (the first

Running head: First direct evidence of two stages in free recall and last items). The middle items are not present in the initial recall at all. The middle items are, however, present in the second process.

The total recall is a simple sum of the items in working memory and recalls from the second stage, there is no additional memory process. Fig. 3 displays the results of the *ab-initio* fitting of the total recall with a linear combination of working memory content and secondary recall (minimizing the sum of squares). The top panels are, from left to right, the 10-2, 15-2 and 20-2 data sets; the bottom panels are, from left to right, the 20-1, 30-1 and 40-1 data sets. The fits are good. Deviations are presumably due to that the initial recall is not a statistical representation of all of working memory for short lists (for very short lists the initial recall will simply consist of the first item). In Table 2 is shown the fitting parameters which provide *ab-initio* values of the capacity of working memory ranging from 3.0 to 3.9 items. The coefficient that indicates working memory capacity increases with the number of items in the list. Perhaps this increase indicates an increased probability of chunking working memory content as the number of items increases. This increase is half of the increase in the coefficient for secondary recall.

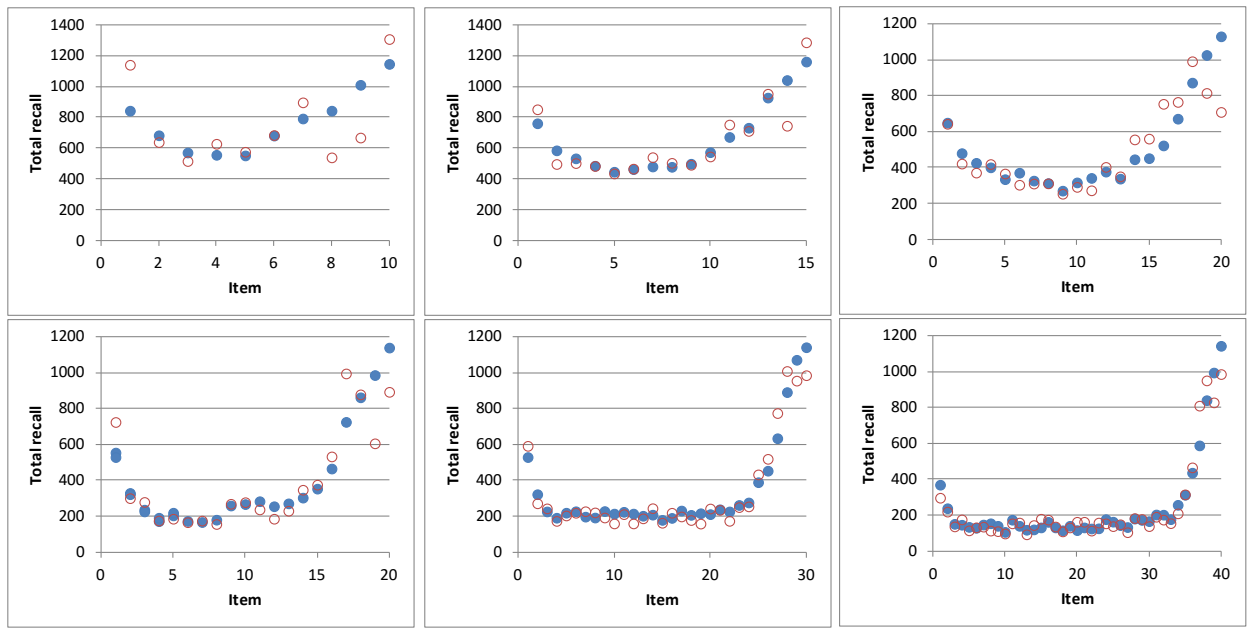

*Fig. 3 Total recall (filled circles) and total recall predicted using the optimal linear combination of the initial and final recalls (unfilled circles). Data sets are 20-1 (left panel), 30-1 (middle panel) and 40-1 (right panel). The coefficients of the linear combinations are in Table 4.*

| Series | 10-2 | 15-2 | 20-2 | 20-1 | 30-1 | 40-1 |
|---|---|---|---|---|---|---|
| Working memory coefficient | 3 | 3.4 (13% increase) | 3.9 (30% increase) | 3.6 | 3.8 (6% increase) | 3.7 (3% increase) |
| Second stage recall coefficient | 3.4 | 4.8 (40% increase) | 4.5 (32% increase) | 3.2 | 4.7 (47% increase) | 4.4 (38% increase) |
| Total recall | 6.4 | 8.2 | 8.4 | 6.8 | 8.5 | 8.1 |
| Rsquared for proportional line | 0.07 | 0.81 | 0.70 | 0.79 | 0.95 | 0.94 |

Table 2. Linear combinations of initial and final recall which provide the best fitting to the total recall and how much of the variance is described by that combination. Numbers in parentheses indicate percent increase over 10-2 (20-1) valuess for the N-2 (N-1) series. Note that both the working memory and second stage recall coefficients increase with list size though the increase is faster for the second stage recall coefficients.

| Series | 10-2 | 15-2 | 20-2 | 20-1 | 30-1 | 40-1 | Murdock & Okada 20 items |
|---|---|---|---|---|---|---|---|
| Lowest TI ratio | 3.61 | 3.60 | 3.81 | 3.06 | 3.65 | 3.11 | 3.15 |
| Item with lowest TI ratio | 10 | 10 | 16 | 17 | 27 | 37 | 17 |

*Table 3. Here is indicated for which items in the Murdock (1962) and Murdock & Okada (1970) datasets the lowest upper limit of the capacity of working memory is derived. The TI ratio refers to the ratio of total recall of a particular item to the initial recall of that item.*

| Series | 10-2 | 15-2 | 20-2 | 20-1 | 30-1 | 40-1 | Average |
|--------|------|------|------|------|------|------|---------|
| Stdev/Average initial | .85 | 1.27 | 1.23 | 1.32 | 1.85 | 2.31 | 1.47 |
| Stdev/Average second stage | .37 | .27 | .43 | .58 | .35 | .46 | 0.41 |

*Table 4. Characteristics of working memory and second stage recall.*

From the requirement that the total recall cannot exceed what is in working memory, item by item, an upper limit on the capacity of working memory can be calculated. This limit is

*For all items i: Working Memory Capacity \* initial recall probability of item i <= total recall probability of item i (Equation 1)*

The ratio of total recall to initial recall of all items (the TI ratio for short), i.e. the upper limit on the capacity of working memory, is shown in Figures 4 and 5 and Table 3. The upper limit of working memory capacity is 3.61, 3.60, 3.81, 3.06, 3.65, and 3.11 for the 10-2, 15-2, 20-2, 20-1, 30-1 and 40-1 data, respectively (for the 40-1 data the numbers corresponding to zero initial recalls were omitted from this calculation since they would give rise to infinities). The Murdock & Okada (1970) data gives us the slightly higher limit of 3.15 from the 17th item out of 20 items. The overall minimum is 3.06 which is then the upper limit of the capacity of working memory[2].

---

[2] The statistical error can be estimated as being less than the sum of independent errors = sqrt(1/(total times item recalled)+1/(initial times item recalled))=sqrt(1/725+1/237)=7% since the two measures are not independent and the corresponding errors would tend to cancel out.

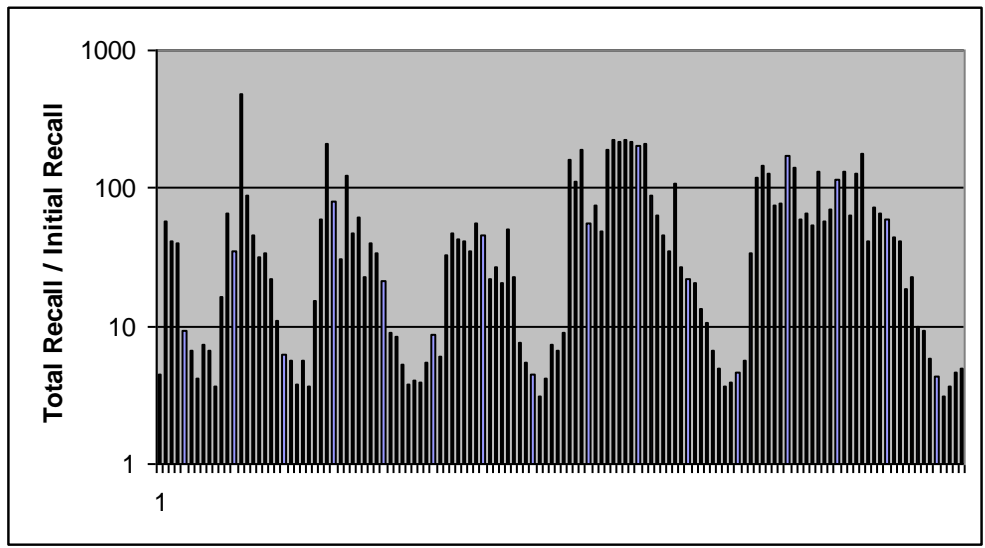

*Figure 4. Upper limits on working memory capacity using the ratio of total recall to initial recall. The points are, in list order, 10-2, 15-2, 20-2, 20-1, 30-1, and 40-1 (with a few items missing from 40-1 in which the total recall is zero). The lowest values, the estimates for the upper limit of the capacity of working memory, tend to come from one of the last word in each series.*

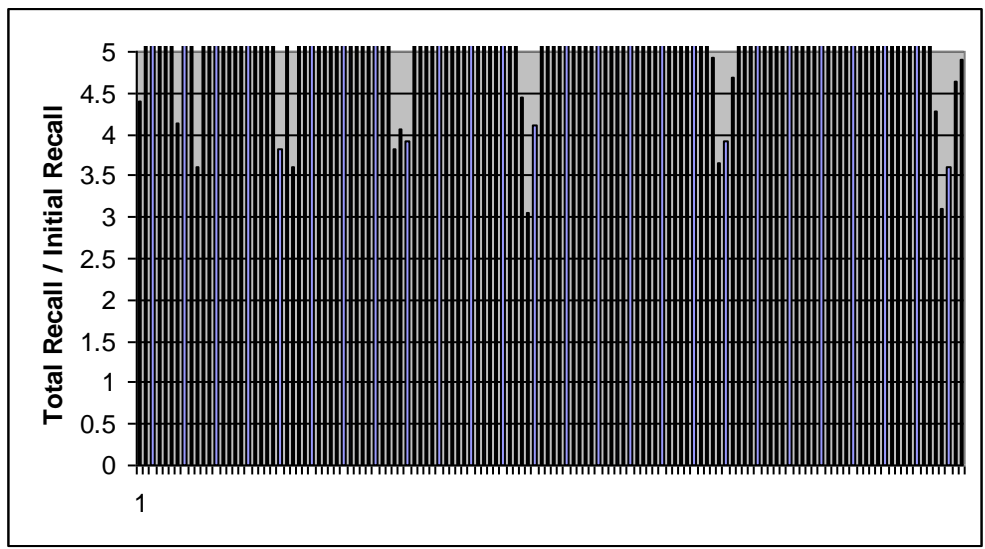

*Figure 5. Upper limits on working memory capacity from Figure 5 with a different scale. The smallest upper limit is 3.06 and comes from the 17th item in the 20-2 series.*

The upper limit that is just larger than 3 is a limit that is in some agreement with recent serial recall experiments (Chen and Cowan, 2009, find a limit of 3) and recent still controversial experimental findings for the capacity of vision working memory (see, for example, Anderson, Vogel and Awh, 2011; for an opposing view see Bays and Husain, 2008; and Bays, Catalao and Husain, 2009). They also agree with the results reviewed in Watkins (1974).

I should note that the limit of 3.06 refers to a statistical average over individuals and to words, not chunks, and, strictly speaking, it refers to words from the Toronto Word Pool.

How quickly is working memory emptied? Using the 40-1 data, Fig. 6 displays the memory content that remains as a function of recall for item 37 (see Table 3), an item which is predominantly in working memory (diamonds). The probability of working memory recall decays exponentially (supporting the conjecture in Murdock, 1962). Fig. 6 also displays the working memory coefficient of the fit of each recall distribution (squares). In both cases it appears that working memory is empty after 4-5 recalls, as was also indicated by the recall distributions in Fig. 1a-b.

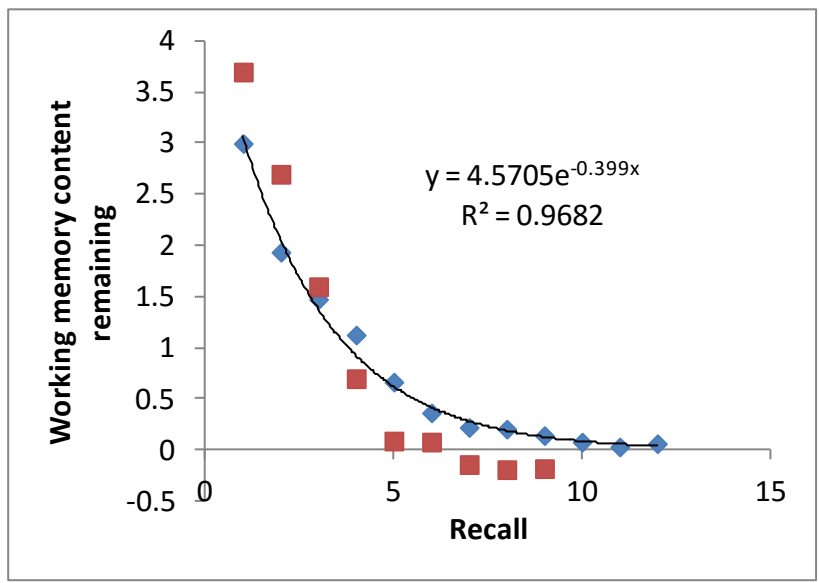

The plotted equation reads:

$$y = 4.5705e^{-0.399x}$$
$$R^2 = 0.9682$$

Fig 6. Initial working memory content remaining in the 40-1 data set calculated by extrapolation from item 37 (diamonds with an exponential decay) and from the coefficient of the linear combination of working memory and second stage recall (squares).

**General Discussion**

The present direct evidence of a two stage model from the separate free recall distributions confirms earlier indirect evidence.  Glanzer & Cunitz (1966) proposed that the serial position curve consists of short-term storage which corresponds to a rising recency effect and a long term storage corresponding to a decline from the beginning to the end and what I find is similar but also different: two <u>stages</u> which presumably come from the same memory store (the same store responsible for short term recognition and cued recall in which content decays in about 15 minutes; Tarnow, 2008 using data from Rubin et al, 1999) in which the second-stage probability of recall of items declines from beginning to the end; working memory, corresponding to Glanzer & Cunitz' short-term storage, is responsible for the recency effect but also has some primacy in it.  A model that describes the current finding almost identically was assumed by Craik (1968):

> "The position adopted here is that after presentation, words are registered both in a limited-capacity primary memory (PM) and also in a much larger secondary memory (SM). Recall is viewed as occurring in two stages: First, a read-out of material still present in PM ; and second, a search process through the relevant part of SM."

Similarly, Murdock (1974) wrote "the subject is assumed to dump primary memory first; that is, the subject outputs those items still in the limited-capacity primary-memory store." (p. 204) I have two quarrels with Craik's articulation.  First, Occam's razor would suggest that both memory sources are the same, the difference is only that items in working memory are kept active while items recalled in the second stage are allowed to decay with time.  Second, when Craik mentions "a much larger secondary memory" he means long term memory in contrast to short term memory; Occam's razor would dictate that short term memory is activated long term memory and I have found elsewhere that this activation decays with a slow logarithm allowing for a very large part of long term memory to remain activated (Tarnow, 2008). The large size of activated long term memory can be seen in cued recall and recognition but not in free recall presumably because the second stage retrieval process cannot find more than a few items in addition to what was kept activated in working memory.

Since working memory should be quantized and presumably consists of exactly three items, my assumption is that there are three nerve bundles or similar corresponding to those three items which keep three items activated.  The representations of those items are the same if the items are activated in the second stage.  I have previously suggested a mechanism for the activation of long term memory (Tarnow, 2009).

The discontinuity in the linear slope (see Fig. 2) shows that there are two and exactly two stages. Previously the search for such discontinuities in word free recall have yielded little. For example, there is no discontinuity in word free recall response times (Tarnow, 2013), nor in errors (Tarnow, 2014) and latency distributions are the same for all but the first recall (Laming, 1999). McElree (2006) showed that there is no discontinuity in item recognition time beyond the first item. Balakrishnanl and Ashby (1992) did not find any discontinuity in reaction time distributions in an experiment asking for enumeration of colored blocks when the number of blocks increases from 1 to 8.

I arrived at the two different types of values of the capacity of working memory. The smallest, 3.06, which has a small error bar, presumably is the value associated with an item that is not chunked. The other values, ranging from 3 to 3.9 for the total recall fit (Table 2) and 4-4.5 for the transition of the slope in Fig. 2 presumably includes chunking of working memory. It is consistent with the values going up as the number of items increases.

Also consistent with the assumption that the initial recall is working memory is that I have shown elsewhere that there is substantial individual control over the initial recall: individual differences are the strongest for the initial recall – see Tarnow, 2015b).

Some forty years ago Murdock (1974) wrote that "we have no satisfactory way to characterize [output-order effects]. (p. 202)" This was accomplished in this article, at least in part. Free recall, while seemingly complex, continuous to elucidate how our memory works, fifty years after the experimental method was introduced.

Running head: First direct evidence of two stages in free recall